\documentclass[10pt,conference]{IEEEtran}
\usepackage{amsfonts}
\usepackage{amsfonts}
\usepackage{setspace}

\usepackage{amsmath}
\usepackage{amssymb}
\usepackage[dvips]{graphicx}
\usepackage{epsfig}
\usepackage{hyperref}

\usepackage{amsmath}
\usepackage{amssymb}
\usepackage[dvips]{graphicx}
\usepackage{epsfig}

\newcommand{\tsnr}{{\text{\footnotesize{SNR}}}}

\newcommand{\E}{\mathbb{E}}

\newcommand{\C}{{\sf{C}}}

\newcommand{\Pb}{\bar{P}}

\newcommand{\figsize}{0.38}

\newcommand{\tMAC}{\text{MAC}}

\newtheorem{Lem1}{Proposition}%[section]
\newtheorem{Lem}{Theorem}

\begin{document}

\title{On the Achievable Throughput Region of Multiple-Access Fading Channels with QoS Constraints}

%\author{\authorblockN{Michael Shell} \and
%\and \authorblockN{James Kirk\\ and Montgomery Scott}
%\authorblockA{Starfleet Academy\\
%San Francisco, California 96678-2391\\ Telephone: (800)
%555--1212\\ Fax: (888) 555--1212}}

% avoiding spaces at the end of the author lines is not a problem with
% conference papers because we don't use \thanks or \IEEEmembership

\author{\authorblockN{Deli Qiao, Mustafa Cenk Gursoy, and Senem
Velipasalar}
\authorblockA{Department of Electrical Engineering\\
University of Nebraska-Lincoln, Lincoln, NE 68588\\ Email:
dqiao726@huskers.unl.edu, gursoy@engr.unl.edu,
velipasa@engr.unl.edu} }

%\specialpapernotice{(Invited Paper)}

\maketitle

\begin{abstract}\footnote{This work was supported by the National Science Foundation under Grants CNS--0834753, and CCF--0917265.}
Effective capacity, which provides the maximum constant arrival
rate that a given service process can support while
satisfying statistical delay constraints, is analyzed in a multiuser scenario. In particular, we study the achievable effective capacity region of
the users in multiaccess fading channels (MAC) in the presence of quality of
service (QoS) constraints. We assume that channel side information
(CSI) is available at both the transmitters and the receiver, and
superposition coding technique with successive decoding is used.
When the power is fixed at the transmitters, we show that varying the decoding
order with respect to the channel state can significantly increase
the achievable \emph{throughput region}. For a two-user case, we
obtain the optimal decoding strategy when the users have the same
QoS constraints. Meanwhile, it is shown that time-division
multiple-access (TDMA) can achieve better performance than
superposition coding with fixed successive decoding order at the
receiver side for certain QoS constraints. For power and rate
adaptation, we determine the optimal power allocation policy with
fixed decoding order at the receiver side. Numerical results are
provided to demonstrate our results.
\end{abstract}

%\begin{spacing}{1.1}
\section{Introduction}
Multiaccess fading channels have been extensively studied over the years from an information-theoretic point of view \cite{gallager}-\cite{suboptdma}. For
instance, Tse and Hanly \cite{polymatroid} have characterized the
capacity region and determined the optimal resource allocation policies. It has
been shown that the boundary surface points are achieved by
successive decoding techniques, and each boundary point is associated
with a weighted maximization of the sum rate. %Although this approach
%is powerful and obtains all boundary points, it is abstract and hard
%to get an intuition about the properties of the solution.
Vishawanath \emph{et al.} \cite{optimumgoldsmith} derived the
explicit optimal power and rate allocation schemes (similar to
\emph{waterfilling}) by considering that the users are successively
decoded in the same order for all channel states. For the convex
capacity region, the unique decoding order was shown to be the
reverse order of the priority weight. Caire \emph{et al.} proved
that TDMA is always suboptimal in low-SNR case \cite{suboptdma}.
On the other hand, these information theoretical studies have not addressed the delay and QoS constraints.

In this paper, we consider statistical QoS constraints and study
the achievable rate region under such constraints in multiaccess fading channels. For this analysis, we employ the
concept of effective capacity \cite{dapeng}, which can be seen as
the maximum constant arrival rate that a given time-varying service
process can support while satisfying statistical QoS guarantees.
Effective capacity formulation uses the large deviations theory and
incorporates the statistical QoS constraints by capturing the rate
of decay of the buffer occupancy probability for large queue
lengths. The analysis and application of effective capacity in
various settings has attracted much interest recently (see e.g.,
\cite{jia}--\cite{liu-cooperation} and references therein). We here consider the
scenario in which both the transmitters and the receiver have the
channel side information (CSI). First, we characterize the rate
regions when the transmitters work at fixed power. Unlike the results
obtained in \cite{gallager}, varying the decoding order is shown to
significantly increase the achievable rate region under QoS
constraints. Also, it is demonstrated that time sharing strategies among the vertex of the
rate regions can no longer achieve the boundary surface. If we take
the sum-rate throughput, or the sum effective capacity, as a
measure, TDMA can even achieve better performance than superposition
coding with fixed decoding order in certain cases. When power
adaptation is considered, we provide the optimal power allocation
policy when the users are being decoded in a fixed order at the receiver
side.
\begin{figure}
\begin{center}
\includegraphics[width=\figsize\textwidth]{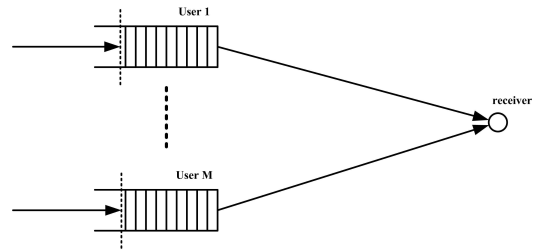}
\caption{The system model.}\label{fig:model}
\end{center}
\end{figure}

The paper is organized as follows. Section II describes the system
model. In Section III, effective capacity as a measure of the performance under
statistical QoS constraints is briefly discussed, and the
\emph{throughput region} under QoS constraints is defined. Section IV includes our main results and presents numerical results. Finally, Section V concludes the paper.

\section{System Model}

As shown in Figure \ref{fig:model}, we consider an uplink scenario
where $M$ users with individual power constraints and QoS
constraints communicate with a single receiver. It is assumed that
the transmitters generate data sequences which are divided into
frames of duration $T$. These data frames are initially stored in
the buffers before they are transmitted over the wireless channel.
The discrete-time signal at the receiver in the
$i^{\text{th}}$ symbol duration is given by
\begin{equation}
Y[i]=\sum_{j=1}^{M}h_j[i]X_j[i]+n[i], \quad i=1,2,\ldots
\end{equation}
where $M$ is the number of users, $X_j[i]$ and $h_j[i]$ denote the
complex-valued channel input and the fading coefficient of the $j$th
user, respectively. We assume that $\{h_j[i]\}$'s are jointly stationary
and ergodic discrete-time processes, and we denote the
magnitude-square of the fading coefficients by $z_j[i]=|h_j[i]|^2$.
The channel input of user $j$ is subject to an average power
constraint $\E\{|x_j[i]|^2\} \le \Pb_j$ for all $j$, and we assume
that the bandwidth available in the system is $B$. $Y[i]$ is the
channel output. Above, $n[i]$ is a zero-mean, circularly symmetric,
complex Gaussian random variable with variance $\E\{|n[i]|^2\} =
N_0$. The additive Gaussian noise samples $\{n[i]\}$ are assumed to
form an independent and identically distributed (i.i.d.) sequence.

\subsection{Fixed Power and Variable Rate}
First, we consider the case in which the transmitters operate at fixed
power.
%when both the receiver and the
%transmitters have the perfect CSI but the transmitters do not adopt
%any power allocation policy.
%For this scenario, we study the effect of the QoS constraints on the
%\emph{throughput region}.
The capacity region of this channel is given by \cite{gallager}:
{\begin{small}
\begin{align}\label{eq:capacityreg}
\mathcal{R}_{\tMAC}&=\Bigg\{\left(R_1,\ldots,R_M\right):\mathbf{R}(S)\leq\nonumber\\
&B\E_{\mathbf{z}}\left\{\log_2\left(1+\sum_{j\in S}\tsnr_j
z_j\right)\right\}\, , \forall S\subset\{1,\ldots,M\}\Bigg\}
\end{align}
\end{small}}
\hspace{-.2cm}where $\tsnr_j=\Pb_j/(N_0 B)$ denotes the average transmitted
signal-to-noise ratio of user $j$, $\mathbf{z}=(z_1,\cdots,z_M)$ is
a random vector comprised of the channel coefficients. As is known,
there are $M!$ vertices for the polyhedron defined in
(\ref{eq:capacityreg}). The vertex
$\mathbf{R}_{\mathbf{\pi}}=\left(R_{\pi(1)},\cdots,R_{\pi(M)}\right)$
corresponds to a permutation $\bf{\pi}$, or the successive decoding
order at the receiver, i.e., users are decoded in the order given by
$\pi(1),\cdots,\pi(M)$. %During the decoding, signals decoded already
%will not interfere with the future decoding, while the signals not
%yet decoded will be viewed as interference.
The vertex is given by :%
%We know that for any fixed decoding order $\mathbf{\pi}$, the
%maximum instantaneous rate for each user is known to be
\begin{small}
\begin{align}\label{eq:caparegvertexrate}
R_{\pi(k)}&=B\E_{\mathbf{z}}\left\{\log_2\left(1+\frac{\tsnr_{\pi(k)}z_{\pi(k)}}{1+\sum_{i=k+1}^{M}\tsnr_{\pi(i)}z_{\pi(i)}}\right)\right\}\nonumber\\
&\phantom{\sum_{i=k+1}^{M}\tsnr_{\pi(i)}z_{\pi(i)}}\quad
\text{bits/s}, \quad k=1,\cdots,M.
\end{align}
\end{small}
which also defines the maximum instantaneous service rate for user
$\pi(k)$ at the given decoding order $\mathbf{\pi}$. Time sharing
among these $M!$ permutations yields any point on the boundary
surface \cite{book}. As can be easily verified, due to the $\log$
term in the expression for the capacity region
(\ref{eq:capacityreg}), varying decoding order according to the
channel state does not provide any improvement for the achievable
capacity region.% But whether the \emph{throughput region} defined
%latter when there exist QoS constraints can be improved by allowing
%the receiver to vary the decoding order with respect to the channel
%state $\mathbf{z}$?
%
%With the successive decoding order $\bf{\pi}$ and the given
%$\textbf{\tsnr}=(\tsnr_1,\cdots,\tsnr_M)$, the instantaneous rate
%for each user is
%\begin{align}\label{eq:instanvertexrate}
%R_{\pi(k)}(\textbf{\tsnr})&=B\log_2\left(1+\tsir_{\pi(k)}z_{\pi(k)}\right)
%\quad \text{bits/s}, \quad k=1,\cdots,M.
%\end{align}
%where
%\begin{align}
%\tsir_{\pi(k)}=\frac{\tsnr_{\pi(k)}z_{\pi(k)}}{1+\sum_{i=k+1}^{M}\tsnr_{\pi(i)}z_{\pi(i)}}
%\end{align}
%represents the instantaneous signal-to-interference ratio (SIR) of
%user $j=\pi(k)$.

\subsection{Variable Power and Variable Rate}
Now, we suppose that dynamic power and rate allocation is performed
according to time-variations in the channels. For a given power allocation
policy $\mathcal {U}=\{\mu_1,\cdots,\mu_M\}$, where
$\mu_j\geq0\,\forall j$ can be viewed as a function of $\mathbf{z}$.
The achievable rates are defined as
\begin{small}
\begin{align}\label{eq:instcapacityregcsit}
\mathcal{R}(\mathcal{U})&=\Bigg\{\mathbf{R}:\mathbf{R}(S)\leq\E_\mathbf{z}\left\{B\log_2\left(1+\sum_{j\in
S}\mu_j(\mathbf{z}) z_j\right)\right\},\, \nonumber\\
&\phantom{\mathbf{R}:\mathbf{R}(S)\leq mathcal{R}(\mathcal{U})}\quad
\forall S\subset \{1,\cdots,M\}\Bigg\}.
\end{align}
\end{small}
The instantaneous rate at a given decoding order can be obtained
similar to (\ref{eq:caparegvertexrate}) with
$\tsnr$ replaced by $\mu$. Then, the rate region is given by
\begin{align}\label{eq:capacityregcsit}
\mathcal{R}_{\tMAC}=\bigcup_{\mathcal{U}\in\mathcal{F}}
\mathcal{R}(\mathcal{U})
\end{align}
where $\mathcal{F}$ is the set of all feasible power control
policies satisfying the average power constraint
\begin{equation}
\mathcal{F}\equiv\{\mathcal{U}:\E_{\mathbf{z}}\left\{\mu_j(\mathbf{z})\leq
\tsnr_j, \mu_j\geq0, \, \forall j\right\}\}
\end{equation}
where $\tsnr_j=\Pb_j/(N_0 B)$ denotes the average transmitted
signal-to-noise ratio of user $j$.

\subsection{TDMA}

For simplicity, we assume that the time division strategy should be
fixed prior to transmission. Let $\delta_j$ denote the fraction of time
allocated to user $j$. Note that we have $\sum_{j=1}^{M}\delta_j=1$. In
each frame, each user occupies the entire bandwidth to transmit the
signal in the corresponding fraction of time. Then, the instantaneous
service rate for user $j$ is given by
\begin{align}\label{eq:tdmarate}
R_{j}(\tsnr_j)=B\log_2\left(1+\frac{\tsnr_j}{\delta_j}z_j\right) \text{ bits/s}
\end{align}

\section{Preliminaries}
\subsection{Effective Capacity}
In \cite{dapeng}, Wu and Negi defined the effective capacity as the
maximum constant arrival rate\footnote{For time-varying arrival
rates, effective capacity specifies the effective bandwidth of the
arrival process that can be supported by the channel.} that a given
service process can support in order to guarantee a statistical QoS
requirement specified by the QoS exponent $\theta$. If we define $Q$
as the stationary queue length, then $\theta$ is the decay rate of
the tail distribution of the queue length $Q$:
\begin{equation}
\lim_{q \to \infty} \frac{\log P(Q \ge q)}{q} = -\theta.
%P\{D(\infty)>D_{\textrm{max}}\}\approx e^{-\theta
%D_{\textrm{max}}}.
\end{equation}
Therefore, for large $q_{\max}$, we have the following approximation
for the buffer violation probability: $P(Q \ge q_{\max}) \approx
e^{-\theta q_{\max}}$. Hence, while larger $\theta$ corresponds to
more strict QoS constraints, smaller $\theta$ implies looser QoS
guarantees. Similarly, if $D$ denotes the steady-state delay
experienced in the buffer, then $P(D \ge d_{\max}) \approx
e^{-\theta \delta d_{\max}}$ for large $d_{\max}$, where $\delta$ is
determined by the arrival and service processes
\cite{tangzhangcross2}.

The effective capacity is given by
\begin{align}\label{eq:effectivedefi}
C(\theta)=-\frac{\Lambda(-\theta)}{\theta}=-\lim_{t\rightarrow\infty}\frac{1}{\theta
t}\log_e{\mathbb{E}\{e^{-\theta S[t]}\}} \,\quad \text{bits/s},
\end{align}
where the expectation is with respect to $S[t] =
\sum_{i=1}^{t}s[i]$, which is the time-accumulated service process.
$\{s[i], i=1,2,\ldots\}$ denote the discrete-time stationary and
ergodic stochastic service process.

In this paper, in order to simplify the analysis while considering
general fading distributions, we assume that the fading coefficients
stay constant over the frame duration $T$ and vary independently for
each frame and each user. In this scenario, $s[i]=T R[i]$, where
$R[i]$ is the instantaneous service rate in the $i$th frame duration
$[iT;(i+1)T]$. Then, (\ref{eq:effectivedefi}) can be written as
\begin{align}
C(\theta)&=%-\lim_{t\rightarrow\infty}\frac{1}{\theta
%Tt}\log_e{\mathbb{E}\{e^{-\theta T \sum_{i=1}^{t}R[i]}\}} \nonumber\\
%&=-\lim_{t\rightarrow\infty}
-\frac{1}{\theta T}\log_e\mathbb{E}_{\mathbf{z}}\{e^{-\theta T
R[i]}\}\,\quad \text{bits/s}, \label{eq:effectivedefirate}
\end{align}
where $R[i]$ denotes the instantaneous rate sequence with respect to
$\mathbf{z}$. (\ref{eq:effectivedefirate}) is obtained using the fact
that instantaneous rates $\{R[i]\}$ vary independently.
 %Obviously, the larger $S[t]$, or $R[i]$
%equivalently, the larger $C(\theta)$.
The effective capacity normalized by bandwidth $B$ is
\begin{equation}\label{eq:normeffectivedefi}
\C(\theta)=\frac{C(\theta)}{B} \quad \text{bits/s/Hz}.
\end{equation}

\subsection{Throughput Region}
Suppose that $\Theta=(\theta_1,\cdots,\theta_M)$ is a vector
composed of the QoS constraints of $M$ users%,
%$\Beta=(\beta_1,\cdots,\beta_M)$, $\beta_j=\frac{\theta_i
%TB}{\log_e2}$ are the normalized QoS constraints
. Let
$\mathbf{\C}(\Theta)=\left(\C_1(\theta_1),\cdots,\C_M(\theta_M)\right)$
denote the vector of the normalized effective capacities. We first
have the following characterization.
\begin{Lem1}\label{prop:throughputreg}
The instantaneous \emph{throughput region} can be defined as
\begin{small}
\begin{align}\label{eq:instanthroureg}
&\mathcal{C}_{\tMAC}(\Theta,\textbf{\tsnr})\nonumber\\
&=\Biggl\{\mathbf{\C}(\Theta)\geq\mathbf{0}:
\C_j(\theta_j)\leq-\frac{1}{\theta_j
TB}\log_e\mathbb{E}_{\mathbf{z}}\left\{e^{-\theta T
R_j[i]}\right\},\nonumber\\
&\phantom{mathcal{C}_{\tMAC}(\Theta,\textbf{\tsnr})} \text{subject
to:}\forall\E\{\mathbf{R[i]}\}\in\mathcal{R}_{\tMAC}\Biggr\}.
\end{align}
\end{small}
where $\mathbf{R}[i]=\{R_1[i],R_2[i],\cdots,R_M[i]\}$ represents
vector composed of the instantaneous rate of $M$ users.% Then the \emph{throughput region} for all possible
%$\mathbf{R}$ is
%\begin{equation}
%\mathcal{C}_{\tMAC}(\Theta,\textbf{\tsnr})=\bigcup_{\mathbf{R}}\mathcal{C}_{\mathbf{R}}(\Theta,\textbf{\tsnr}).
%\end{equation}
\end{Lem1}

\emph{Remark: }The \emph{throughput region} defined in Proposition
\ref{prop:throughputreg} represents the set of all vectors of
constant arrival rates such that there exists a possible
instantaneous rate adaptation $\mathbf{R}[i]$ among the $M$ users,
which can guarantee the QoS constraints
$\Theta=(\theta_1,\cdots,\theta_M)$.

\emph{Corollary: }The \emph{throughput region} for TDMA can be
deemed as the achievable vectors of arrival rates with each
component bounded by the effective capacity obtained for the instantaneous
service rate given in (\ref{eq:tdmarate}). The effective capacity for user
$j$ on the boundary surface becomes
\begin{align}
\hspace{-.2cm}\C^{\text{TD}}_{j}(\theta_j)=-\frac{1}{\theta_j TB}\log_e
\E\left\{e^{-\delta_j \theta_j
TB\log_2\left(1+\frac{\tsnr_j}{\delta_j}z_j\right)}\right\}
\end{align}
We assume that $\E\{\mathbf{R}[i]\}$ can take any possible values
defined in the $\mathcal{R}_{\tMAC}$. We  have the following premilinary
result.
\begin{Lem}\label{theo:convexset}
The \emph{throughput region}
$\mathcal{C}_{\tMAC}(\Theta,\textbf{\tsnr})$ is convex.
\end{Lem}
\emph{Proof: }Let $\C_1(\Theta)$ and $\C_2(\Theta)$ belong to
$\mathcal{C}_{\tMAC}(\Theta,\textbf{\tsnr})$. Therefore, there
exists some $\mathbf{R}[i]$ and $\mathbf{R'}[i]$ for $\C_1(\Theta)$
and $\C_2(\Theta)$, respectively. By a time sharing strategy, for
any $\alpha\in(0,1)$, we know that
$\E\{\alpha\mathbf{R}[i]+(1-\alpha)\mathbf{R'}[i]\}\in\mathcal{R}_{\tMAC}$.
\begin{align}
&\alpha\C_1+(1-\alpha)\C_2\nonumber\\
&=-\frac{1}{\Theta TB}\log_e\left(\E\left\{e^{-\Theta
T\mathbf{R}[i]}\right\}\right)^\alpha\left(\E\left\{e^{-\Theta
T\mathbf{R'}[i]}\right\}\right)^{1-\alpha}\nonumber\\
&=-\frac{1}{\Theta TB}\log_e\left(\E\left\{\left(e^{-\Theta
T\alpha\mathbf{R}[i]}\right)^{\frac{1}{\alpha}}\right\}\right)^\alpha\nonumber\\
&\phantom{-\frac{1}{\Theta TB}e^{-\Theta
T\mathbf{R'}}}\cdot\left(\E\left\{\left(e^{-\Theta
T(1-\alpha)\mathbf{R'}[i]}\right)^{\frac{1}{1-\alpha}}\right\}\right)^{1-\alpha}\nonumber\\
&\leq-\frac{1}{\Theta TB}\log_e\E\left\{e^{-\Theta T\left(\alpha
\mathbf{R}[i]+(1-\alpha)\mathbf{R'}[i]\right)}\right\}
\end{align}
where the vector operation is with respect to each component, and
Holder's inequality is used. Hence, $\alpha\C_1+(1-\alpha)\C_2$
still lies in the \emph{throughput region}. \hfill$\square$

We are interested in the boundary of the region
$\mathcal{C}_{\tMAC}(\Theta,\textbf{\tsnr})$. %The boundary surface
%is the set of those rates such that no component can be increased
%with the other components fixed, while remaining in the
%\emph{throughput region} \cite{polymatroid}.
Now that
$\mathcal{C}_{\tMAC}(\Theta,\textbf{\tsnr})$ is convex, we can
characterize the boundary surface by considering the following
optimization problem \cite{polymatroid}:
\begin{equation}\label{eq:maxproblem}
\max \mathbf{\lambda}\cdot \C(\Theta) \quad \text{subject to:}
\C(\Theta)\in \mathcal{C}_{\tMAC}(\Theta,\textbf{\tsnr}).
\end{equation}
for all priority vectors
$\mathbf{\lambda}=(\lambda_1,\cdots,\lambda_M)$ in
$\mathfrak{R}^{M}_{+}$ with $\sum_{j=1}^{M}\lambda_j=1$.

\section{Multiple-Access Channels with QoS Constraints}
%In this Section, we first assume that the signals are transmitted at
%a constant power level for each frame. Then, with any fixed decoding
%order, we try to find the optimal power control policy.

\subsection{MAC without Power Control}
If we assume that the receiver decodes the users at a fixed order,
it is obvious that only the vertices can be achievable. Suppose that
time sharing technique is employed. Moreover, assume that the time fraction
for each order $\pi_{m}$ is $\tau_m$, such that $\tau_m\geq0$ and
$\sum_{m=1}^{M!}\tau_m=1$. Then, the effective capacity for each
user is
\begin{align}
\C_j(\theta_j)=-\frac{1}{\theta_j
TB}\log_e\E_{\mathbf{z}}\left\{e^{-\theta_j T\sum_{m=1}^{M!}\tau_m
R_{\pi^{-1}_m(j)}}\right\}
\end{align}
where $ R_{\pi^{-1}_m(j)}[j]$ represents the instantaneous service
rate of user $j$ at a given decoding order $\pi_{m}$, which is given
by
\begin{align}
 R_{\pi^{-1}_m(j)}=B\log_2\left(1+\frac{\tsnr_j
z_j}{1+\sum_{\pi^{-1}_m(i)>\pi^{-1}_m(j)}\tsnr_i z_i}\right)
\end{align}
where $\pi^{-1}_m$ is the inverse trace function of $\pi_{m}$.% which
%defines the position of $j$ in the order $\pi_{m}$.

If the receiver has the freedom to change the decoding order
according to the estimated channel state, we suppose there exists a
rate allocation policy $\mathbf{R}[i]$ for any
$\lambda\in\mathfrak{R}^M_+$. In this paper, we consider a class of
successive decoding techniques $\mathbf{\mathcal {F}}(\mathbf{z})$
parameterized as a function of the channel states $\mathbf{z}$. More
specifically, the vector space $\mathfrak{R}^{M}_{+}$ for
$\mathbf{z}$ is divided into disjoint $\mathcal{Z}_m,
m\in\{1,2,\ldots,M!\}$ regions with respect to each
$\pi_m$\footnote{Each region corresponds to a unique $\pi$.}. For
instance, when $\mathbf{z}\in\mathcal{Z}_1$, the base station will
decode the information in the order $M,M-1,\ldots,1$. Then, the effective
capacity for each user is
\begin{align}
\C_j(\theta_j)&=-\frac{1}{\theta_j
TB}\log_e\E_{\mathbf{z}}\left\{e^{-\theta_j T R_j}\right\}\nonumber\\
&=-\frac{1}{\theta_j TB}\log_e\Bigg(\sum_{m=1}^{M!}
\int_{\mathbf{z}\in\mathcal{Z}_m}e^{-\theta_j T
R_{\pi^{-1}_m(j)}}d\mathbf{z}\Bigg)
\end{align}
Considering the expression for effective capacity and the
optimization problem in (\ref{eq:maxproblem}), the optimal rate
adaptation with respect to the channel state seems intractable. In this paper, we consider a simplified scenario in which all users
have the same QoS constraint described by $\theta$. This case arises, for instance, if users do not have priorities
over others in terms of buffer limitations or delay constraints.
% Previous works on
%guaranteeing stable queues proposed some useful scheduling methods,
%such as Longest-Queue-Highest-Possible-Rate (LQHPR)\cite{lqhpr},
%exponential rule (EXP) \cite{exp}.
\subsubsection{Two-user MAC}
%First, we consider the two-user MAC case.
%We suppose that the two users have the same QoS exponent $\theta$.
Similar to the discussion in \cite{scheduling}, finding an optimal
scheduling scheme can be reduced to finding a function $z_2=g(z_1)$
in the state space such that users are decoded in the order 1,2 if
$z_2<g(z_1)$ and users are decoded in the order 2,1 if $z_2>g(z_1)$.
The problem in (\ref{eq:maxproblem}) becomes
\begin{align}
\max \lambda_1\C_1(\theta,g(z_1))+(1-\lambda_1)\C_2(\theta,g(z_1))
\end{align}
where $\C_1(\theta,g(z_1))$ and $\C_2(\theta,g(z_1))$ are expressed
in (\ref{eq:capacityrep1}) and (\ref{eq:capacityrep2}) at the top of the next
page. Implicitly, $g(z_1)$ should always be larger than 0 in
(\ref{eq:capacityrep1}) and (\ref{eq:capacityrep2}) in order for the
integral to hold, which may not be guaranteed due to the complexity
of the problem. In that case, we need to find a function
$z_1=g(z_2)$ instead, as will be indicated later.
\begin{figure*}
\begin{align}\label{eq:capacityrep1}
\C_1(\theta,g(z_1))&=-\frac{1}{\theta TB}\log_e\Bigg(
\int_0^{\infty}\int^{\infty}_{g(z_1)}e^{-\theta T
B\log_2\left(1+\tsnr_1z_1\right)}p_{z_2}(z_2)p_{z_1}(z_1)dz_2dz_1\nonumber\\
&\quad\quad
+\int_0^{\infty}\int_{0}^{g(z_1)}e^{-\theta T
B\log_2\left(1+\frac{\tsnr_1z_1}{1+\tsnr_2z_2}\right)}p_{z_2}(z_2)p_{z_1}(z_1)dz_2dz_1\Bigg)
\end{align}
\begin{align}\label{eq:capacityrep2}
\C_2(\theta,g(z_1))&=-\frac{1}{\theta
TB}\log_e\Bigg(\int_0^{\infty}\int_{0}^{g(z_1)}e^{-\theta T
B\log_2\left(1+\tsnr_2z_2\right)}p_{z_2}(z_2)p_{z_1}(z_1)dz_2dz_1\nonumber\\
&\quad\quad+ \int_0^{\infty}\int^{\infty}_{g(z_1)}e^{-\theta T
B\log_2\left(1+\frac{\tsnr_2z_2}{1+\tsnr_1z_1}\right)}p_{z_2}(z_2)p_{z_1}(z_1)dz_2dz_1\Bigg)
\end{align}
\hrule
\end{figure*}
\begin{Lem1}
The optimal scheduling scheme for a specific common QoS constraint
$\theta$ in the two-user case is given by
\begin{align}
g(z_1)&= \frac{(1+\tsnr_1z_1)K^{\frac{1}{\beta}}-1}{\tsnr_2},\,\quad
K\in[1,\infty)\label{eq:scheduresult1}\\
g(z_2)&=
\frac{(1+\tsnr_2z_2)K^{-\frac{1}{\beta}}-1}{\tsnr_1},\,\quad
K\in[0,1)\label{eq:scheduresult2}
\end{align}
where $\beta=\frac{\theta TB}{\log_e2}$, $K\in [0,\infty)$ is some
constant.
\end{Lem1}
\emph{Proof: }Suppose that the optimal scheduling is given by
$z_2=g(z_1)$. We denote
\begin{align}
\mathcal
{J}(g_1(z_1))=\lambda_1\C_1(\theta,g_1(z_1))+(1-\lambda_1)\C_2(\theta,g_1(z_1))
\end{align}
where $g_1(z_1)=g(z_1)+s\eta(z_1)$. $g(z_1)$ is the optimal
scheduling function, $s$ is any constant, and $\eta(z_1)$ represents
arbitrary variation. A necessary condition that needs to be
satisfied is \cite{physics}
\begin{equation}\label{eq:scheducond}
\frac{d}{ds}\left(\mathcal{J}(g_1(z_1))\right)\bigg|_{s=0}=0.
\end{equation}
Define the following (for i=1,2):
\begin{small}
\begin{align}\label{eq:scheduproof1}
\phi_1&=\int_0^{\infty}\int^{\infty}_{g(z_1)}e^{-\theta T
B\log_2\left(1+\tsnr_1z_1\right)}p_{z_2}(z_2)p_{z_1}(z_1)dz_2dz_1\nonumber\\
&\quad +\int_0^{\infty}\int_{0}^{g(z_1)}e^{-\theta T
B\log_2\left(1+\frac{\tsnr_1z_1}{1+\tsnr_2z_2}\right)}p_{z_2}(z_2)p_{z_1}(z_1)dz_2dz_1
\end{align}
\begin{align}\label{eq:scheduproof2}
\phi_2&=\int_0^{\infty}\int_{0}^{g(z_1)}e^{-\theta T
B\log_2\left(1+\tsnr_2z_2\right)}p_{z_2}(z_2)p_{z_1}(z_1)dz_2dz_1\nonumber\\
&\quad\quad+ \int_0^{\infty}\int^{\infty}_{g(z_1)}e^{-\theta T
B\log_2\left(1+\frac{\tsnr_2z_2}{1+\tsnr_1z_1}\right)}p_{z_2}(z_2)p_{z_1}(z_1)dz_2dz_1
\end{align}
\end{small}
\begin{figure}
\begin{center}
\includegraphics[width=\figsize\textwidth]{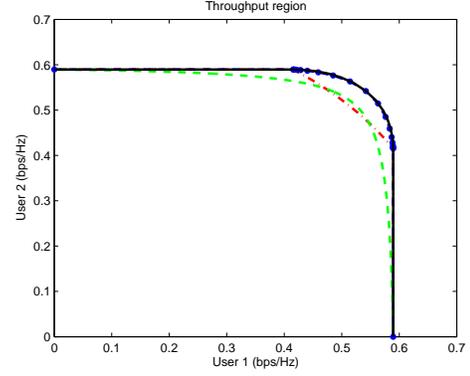}
\caption{The throughput region of two-user MAC case.
$\tsnr_1=\tsnr_2=0$ dB. $\theta_1=\theta_2=0.01$. The solid,
dot-dashed, dashed and dotted lines represent the regions achieved
by optimal scheduling, suboptimal scheduling, fixed decoding with
time sharing, and the TDMA respectively. }\label{fig:fixedsnr}
\end{center}
\end{figure}
By noting that $\frac{dg_1(z_1)}{ds}=\eta(z_1)$, and from
(\ref{eq:scheducond})-(\ref{eq:scheduproof2}), we can derive
\begin{small}
\begin{align}
&\int_0^{\infty}\Bigg(-\frac{\lambda_1}{\theta
TB\phi_1}\left(\left(1+\frac{\tsnr_1z_1}{1+\tsnr_2g(z_1)}\right)^{-\beta}-\left(1+\tsnr_1z_1\right)^{-\beta}\right)\nonumber\\
&\quad-\frac{1-\lambda_1}{\theta
TB\phi_2}\left(\left(1+\tsnr_2g(z_1)\right)^{-\beta}-\left(1+\frac{\tsnr_2g(z_1)}{1+\tsnr_1z_1}\right)^{-\beta}\right)\Bigg)\nonumber\\
&\quad\quad \cdot p_{z_2}(g(z_1))p_{z_1}(z_1)\eta(z_1)dz_1=0
\end{align}
\end{small}
Since the above equation holds for any $\eta(z_1)$, it follows that
\begin{small}
\begin{align}
&-\frac{\lambda_1}{\theta
TB\phi_1}\left(\left(1+\frac{\tsnr_1z_1}{1+\tsnr_2g(z_1)}\right)^{-\beta}-\left(1+\tsnr_1z_1\right)^{-\beta}\right)\nonumber\\
&-\frac{1-\lambda_1}{\theta
TB\phi_2}\left(\left(1+\tsnr_2g(z_1)\right)^{-\beta}-\left(1+\frac{\tsnr_2g(z_1)}{1+\tsnr_1z_1}\right)^{-\beta}\right)=0
\end{align}
\end{small}
which after rearranging and defining $K$ as follows yields
\begin{small}
\begin{align}\label{eq:schedufinal1}
\frac{\left(1+\frac{\tsnr_1z_1}{1+\tsnr_2g(z_1)}\right)^{-\beta}-\left(1+\tsnr_1z_1\right)^{-\beta}}{\left(1+\frac{\tsnr_2g(z_1)}{1+\tsnr_1z_1}\right)^{-\beta}-\left(1+\tsnr_2g(z_1)\right)^{-\beta}}=\frac{(1-\lambda_1)\phi_1}{\lambda_1\phi_2}=K.
\end{align}
\end{small}
Obviously, $K\geq0$. Notice that after simple computation, (\ref{eq:schedufinal1}) % can also be
%written as (\ref{eq:schedufinal2}) in the next page.
%\begin{figure*}
%\begin{align}\label{eq:schedufinal2}
%\frac{(1+\tsnr_1z_1)^{-\beta}\left(\left(1+\tsnr_1z_1+\tsnr_2g(z_1)\right)^{-\beta}-\left(1+\tsnr_1z_1\right)^{-\beta}\left(1+\tsnr_2g(z_1)\right)^{-\beta}\right)}
%{(1+\tsnr_2g(z_1))^{-\beta}\left(\left(1+\tsnr_1z_1+\tsnr_2g(z_1)\right)^{-\beta}-\left(1+\tsnr_2g(z_1)\right)^{-\beta}\left(1+\tsnr_1z_1\right)^{-\beta}\right)}=K
%\end{align}
%\hrule
%\end{figure*}
%which further
becomes
\begin{equation}\label{eq:scheduderiv}
\left(\frac{1+\tsnr_1z_1}{1+\tsnr_2g(z_1)}\right)^{-\beta}=K
\end{equation}
which is (\ref{eq:scheduresult1}). Note here that if $K<1$,
$g(z_1)<0$ for $z_1<\frac{K^{-\frac{1}{\beta}}-1}{\tsnr_1}$. Then
the expressions in (\ref{eq:capacityrep1}) and
(\ref{eq:capacityrep2}) cannot hold. In this case, we denote the
optimal scheduling as $z_1=g(z_2)$ instead. Following a similar approach
as shown from (\ref{eq:capacityrep1})-(\ref{eq:scheduderiv}) will
give us (\ref{eq:scheduresult2}). \hfill$\square$

%In the derivation of the optimal scheduling scheme, although
%implicity, $K=1$ can be achieved with $\lambda_1=1-\lambda_1$, which
%is when the maximum sum-rate throughput can be achieved. With $K=1$,
%the maximum sum-rate can be expressed as the sum of the effective
%capacity for different users.

\subsubsection{Suboptimal Scheduling}
When all users have the same QoS constraint specified by $\theta$, we propose a
suboptimal decoding order given by
\begin{align}
\frac{\lambda_{\pi(1)}}{z_{\pi(1)}}\leq\frac{\lambda_{\pi(2)}}{z_{\pi(2)}}\cdots
\leq\frac{\lambda_{\pi(M)}}{z_{\pi(M)}},
\end{align}
due to the observation that whichever $\lambda_j$ approaches 1,
it should be decoded last.
Considering a two-user example, we can
express the points on the boundary surface as
\begin{small}
\begin{align}
\C_1(\theta)&=-\frac{1}{\theta TB}\log_e\Bigg(
\int_0^{\infty}\int^{\infty}_{\frac{\lambda_2z_1}{\lambda_1}}e^{-\theta
T
B\log_2\left(1+\tsnr_1z_1\right)}dz_2dz_1\nonumber\\
&\quad\quad+\int_0^{\infty}\int_{0}^{\frac{\lambda_2z_1}{\lambda_1}}e^{-\theta
T
B\log_2\left(1+\frac{\tsnr_1z_1}{1+\tsnr_2z_2}\right)}dz_2dz_1\Bigg)
\end{align}
\begin{align}
\C_2(\theta)&=-\frac{1}{\theta
TB}\log_e\Bigg(\int_0^{\infty}\int_{0}^{\frac{\lambda_2z_1}{\lambda_1}}e^{-\theta
T
B\log_2\left(1+\tsnr_2z_2\right)}dz_2dz_1\nonumber\\
&\quad\quad+
\int_0^{\infty}\int^{\infty}_{\frac{\lambda_2z_1}{\lambda_1}}e^{-\theta
T
B\log_2\left(1+\frac{\tsnr_2z_2}{1+\tsnr_1z_1}\right)}dz_2dz_1\Bigg).
\end{align}
\end{small}
%According to the numerical result, the rate region derived by this
%suboptimal scheme can almost perfectly match the one by the optimal
%scheduling scheme.

\begin{figure}
\begin{center}
\includegraphics[width=\figsize\textwidth]{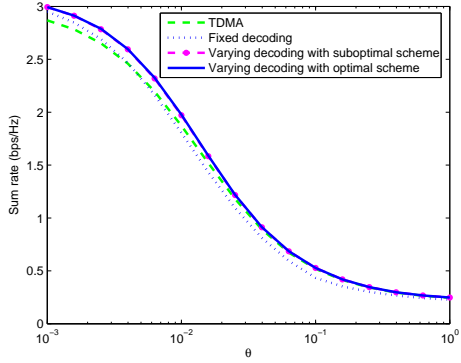}
\caption{The sum-rate throughput as a function of $\theta$.
$\tsnr_1=10$ dB; $\tsnr_2=0$ dB.}\label{fig:comparisonsumrate}
\end{center}
\end{figure}

We have performed numerical analysis over Rayleigh fading channels with
$\E\{\mathbf{z}\}=\mathbf{1}$. In Fig. \ref{fig:fixedsnr} where the throughput region of a two-user MAC is plotted,
we observe that varying the decoding order can significantly increase
the achievable rate region. Moreover, we see that the suboptimal strategy can
achieve almost the same rate region as the optimal strategy. This
can be attributed to the fact that with the optimal strategy, the
receiver can choose the decoding order according to the channel
state such that the weighted sum of effective capacities, i.e.,
summation of $\log$-moment generate functions, is maximized.
Meanwhile, TDMA can achieve some points outside of the
\emph{throughput region} with fixed decoding order at the receiver
side. If sum-rate throughput, i.e. the summation of
the effective capacities, is considered,  we note in Fig.
\ref{fig:comparisonsumrate} that as $\theta$ increases, the curves
of different strategies converge, and as $\theta$ approaches to 0,
TDMA again becomes suboptimal. This may be in large due to the fact
that the transmitted energy is concentrated in the corresponding
fraction of time for each user, which will introduce considerable
weighted sum of throughput as QoS constraints become more stringent,
i.e., the supported throughput becomes smaller. As QoS constraints approach 0, this phenomenon
can be nicely captured by previous work on the Shannon ergodic capacity. %Although we cannot
%find the specific QoS constraints that TDMA perform better, we can
%argue that as QoS constraints become more stringent, the power sharp
%effect can introduce more benefit, achieving larger sum-rate
%throughput.

%
%\begin{figure}
%\begin{center}
%\includegraphics[width=\figsize\textwidth]{comparisonofsumrate.eps}
%\caption{The gain in sum-rate throughput from varying decoding order
%as a function of $\theta_2$ with $\theta_1=0.01$.
%$\tsnr_1=\tsnr_2=0$ dB.}\label{fig:comparisonofsumrate}
%\end{center}
%\end{figure}
\vspace{-.2cm}
\subsection{MAC with Power Control}

In this part, we consider the power control policies with fixed
decoding order at the receiver side. Due to the convexity of
$\mathcal{C}_{\tMAC}$, there exist Lagrange multipliers
$\mathbf{\kappa}\in\mathfrak{R}^M_+$ such that $\C^*(\Theta)$ on the
boundary surface is a solution to the optimization problem
\begin{align}\label{eq:maxproblemcsit}
\max_{\mathbf{\mu}} \mathbf{\lambda}\cdot
\C(\Theta)+\kappa\cdot\E\{\mu\}.
\end{align}
For a given permutation $\pi$, $\C_j(\theta_j)$ is
given by
\begin{align}
\hspace{-.3cm}\C_j(\theta_j)=-\frac{1}{\theta_j TB}\log_e
\E\left\{e^{-\theta_j TB\log_2\left(1+\frac{\mu_j
z_j}{1+\sum_{\pi^{-1}(i)>\pi^{-1}(j)}\mu_i z_i}\right)}\right\}.
\end{align}
Now, the optimization problem (\ref{eq:maxproblemcsit}) is
equivalent to
\begin{align}\label{eq:optiproblemeff}
&\max_{\mu}\sum_{j=1}^{M}-\lambda_j\frac{1}{\theta_j TB}\log_e
\E\left\{e^{-\theta_j TB\log_2\left(1+\frac{\mu_j
z_j}{1+\sum_{\pi^{-1}(i)>\pi^{-1}(j)}\mu_i
z_i}\right)}\right\}\nonumber\\
&\phantom{\max_{\mu}\sum_{j=1}^{M}-\lambda_j\theta_j
TB}+\sum_{j=1}^M\kappa_j\E\{\mu_j\}.
\end{align}
%Note that the above problem is an explicit function of
%$\mu_j\,\forall j$, and since the decoding order is fixed for all
%states, $\mu_j$ will not appear we can concentrate on solving for
%$\mu_j \, \forall j$.
Note that with a fixed decoding order, the user $\pi(M)$ sees no
interference from the other users, and hence the derivative of
(\ref{eq:optiproblemeff}) with respect to $\mu_{\pi(M)}$ will only
be related to the effective capacity formulation of user $\pi(M)$.
Therefore, we can solve an equivalent problem by maximizing $\C_{\pi(M)}$
instead. After we derive $\mu_{\pi(M)}$, the derivative of
(\ref{eq:optiproblemeff}) with respect to $\mu_{\pi(M-1)}$ will only
be related to the effective capacity formulation of user $\pi(M-1)$.
By repeated application of this procedure, with given $\lambda$, (\ref{eq:optiproblemeff})
can be further decomposed into the following $M$ sequential
optimization problems
\begin{align}
&\max_{\mu}-\lambda_j\frac{1}{\theta_j TB}\log_e
\E\left\{e^{-\theta_j TB\log_2\left(1+\frac{\mu_j
z_j}{1+\sum_{\pi^{-1}(i)>\pi^{-1}(j)}\mu_i
z_i}\right)}\right\}\nonumber\\
&\phantom{\max_{\mu}\sum_{j=1}^{M}-\lambda_j\theta_j
TB}+\kappa_j\E\{\mu_j\}\, \quad j\in\{1,\cdots,M\}.
\end{align}
in the inverse order of $\pi$. Similar to \cite{jia}, solving the
above $M$ parallel optimizations is the same as solving
\begin{align}\label{eq:optiproblemeffj}
&\min_{\mu}\E\left\{e^{-\theta_j TB\log_2\left(1+\frac{\mu_j
z_j}{1+\sum_{\pi^{-1}(i)>\pi^{-1}(j)}\mu_i
z_i}\right)}\right\}\nonumber\\
&\phantom{\max_{\mu}\sum_{j=1}^{M}-\lambda_j\theta_j
TB}+\kappa_j\E\{\mu_j\}\, \quad j\in\{1,\cdots,M\}.
\end{align}
Differentiating the Lagrangians with respect to $\mu_j$ respectively
and setting the derivatives to zero yields
\begin{small}
\begin{align}\label{eq:optipowerresult}
\mu_j&=\Bigg(\frac{\left(1+\sum_{ \pi^{-1}(i) > \pi^{-1}(j) } \mu_i
z_i\right)^{\frac{\beta_j}{\beta_j+1}}}{\alpha_j^{\frac{1}{\beta_j+1}}z_j^{\frac{\beta_j}{\beta_j+1}}}\nonumber\\
&\phantom{\pi^{-1}(i) >
\pi^{-1}(j)}-\frac{1+\sum_{\pi^{-1}(i)>\pi^{-1}(j)}\mu_i
z_i}{z_j}\Bigg)^+
\end{align}
\end{small}
where $\beta_j=\frac{\theta_j TB}{\log_e2}$ is the normalized QoS
exponent, $(x)^+=\max\{x,0\}$ and $(\alpha_1,\cdots,\alpha_M)$
satisfy the average power constraints.
%\begin{align}
%\E\Bigg\{&\Bigg(\frac{1}{\alpha_j^{\frac{1}{\beta_j+1}}\left(\frac{z_j}{1+\sum_{\pi^{-1}(i)>\pi^{-1}(j)}\mu_i
%z_i}\right)^{\frac{\beta_j}{\beta_j+1}}}\nonumber\\
%&\quad\quad-\frac{1}{\frac{z_j}{1+\sum_{\pi^{-1}(i)>\pi^{-1}(j)}\mu_i
%z_i}}\Bigg)^+\Bigg\}=\tsnr_j.
%\end{align}
Exploiting the result in (\ref{eq:optipowerresult}), we can find
that instead of adapting power according to its channel state as in
\cite{jia}, the user adapts power according to its channel state
normalized by the interference and the noise observed. Depending on
whether each user is transmitting or not, the vector space
$\mathfrak{R}^M_+$ for $\mathbf{z}$ can be divided into $2^M$
disjoint regions $\mathcal{Z}_m, m\in\{1,\cdots,2^M\}$.

To give an explicit idea of the power control policy, we consider a
two-user example where the decoding order is given by $2,1$. For this case, we
have
\begin{small}
\begin{align}
\mu_1&=\left\{\begin{array}{ll}
\frac{1}{\alpha_1^{\frac{1}{\beta_1+1}}z_1^{\frac{\beta_1}{\beta_1+1}}}-\frac{1}{z_1}
& z_1>\alpha_1,\\
0 & \text{otherwise.}
\end{array}\right.
\intertext{and} \mu_2&=\left\{\begin{array}{ll}
\frac{1}{\alpha_2^{\frac{1}{\beta_2+1}}z_2^{\frac{\beta_2}{\beta_2+1}}}-\frac{1}{z_2}
& z_1\leq\alpha_1 \& z_2>\alpha_2, \\
\frac{\left(\frac{z_1}{\alpha_1}\right)^{\frac{\beta_2}{(\beta_1+1)(\beta_2+1)}}}{\alpha_2^{\frac{1}{\beta_2+1}}z_2^{\frac{\beta_2}{\beta_2+1}}}-\frac{\left(\frac{z_1}{\alpha_1}\right)^{\frac{1}{\beta_1+1}}}{z_2}
& {z_1>\alpha_1 \&
\frac{z_2}{\alpha_2}>\left(\frac{z_1}{\alpha_1}\right)^{\frac{1}{\beta_1+1}}}\\
0 & \text{otherwise}.
\end{array}\right.
\end{align}
\end{small}
where $\alpha_1$ and $\alpha_2$ are chosen to satisfy the average
power constraints of the two users.

\section{Conclusion}
In this paper, we have studied the achievable rate regions in
multi-access fading channels when users operate under QoS constraints. %We have defined
%the \emph{throughput region} while satisfying the statistical delay
%constraints.
With the assumption that both the transmitters and the
receiver have CSI, we have considered different scenarios under which we have
investigated the achievable rate regions. Without power control,
varying the decoding order is shown to significantly increase the achievable rate
region. We have also shown that TDMA can perform better than superposition coding with
fixed decoding
order for certain QoS constraints. %And as the QoS
%constraints become more stringent, TDMA appears to the optimal
%transmission strategy.
For a two-user case with same QoS constraints, the optimal strategy
for varying decoding order is derived, and a simpler suboptimal
decoding rule is proposed which can almost perfectly match the
optimal \emph{throughput region}. Numerical results are provided as
well. Furthermore, we have derived the optimal power control
policies for any given fixed decoding order. %In general, a study on
%the achievable rate regions in MAC channels with QoS constraints has
%been successively carried out.

%
%\end{spacing}

\end{document}